\documentstyle[12pt,epsf,epsfig,here]{article}
\def\beq{\begin{equation}}
\def\eeq{\end{equation}}
\def\bea{\begin{eqnarray}}
\def\eea{\end{eqnarray}}
\def\bq{\begin{quote}}
\def\eq{\end{quote}}

\def\CJP{{\it Can.J.Phys.} }

\def\JMP{{\it J.Math.Phys.} }

\def\JCP{{\it J.Chem.Phys.} }

\def\PR{{\it Phys.Rev.} }
\def\PRL{{\it Phys.Rev.Lett.} }
\def\PRS{{\it Proc.Roy.Soc.} }

\def\ZP{{\it Z.Phys.} }

\def\gappeq{\mathrel{\rlap {\raise.5ex\hbox{$>$}}
{\lower.5ex\hbox{$\sim$}}}}

\def\lappeq{\mathrel{\rlap{\raise.5ex\hbox{$<$}}
{\lower.5ex\hbox{$\sim$}}}}

\begin{document}
\pagestyle{empty}
\begin{flushright}
{CERN-TH/98-23}\\
\end{flushright}
\vspace*{5mm}
\begin{center}
{\bf STABILITY OF THREE- AND FOUR-BODY COULOMB SYSTEMS}
\\
\vspace*{1cm}
{\bf Andr\'e MARTIN} \\
\vspace{0.3cm}
Theoretical Physics Division, CERN \\
CH - 1211 Geneva 23 \\
and
\\
LAPP - F 74941 ANNECY LE VIEUX\\
e-mail: {\tt martina@mail.cern.ch}\\
\vspace*{2cm}
{\bf ABSTRACT} \\ \end{center}
\vspace*{5mm}

We discuss the stability of three- and four-particle system interacting by
pure Coulomb interactions, as a function of the masses and charges of the
particles. We present a certain number of general properties which allow
to answer a certain number of questions without or with less numerical
calculations.

\vspace*{1cm}

\begin{center}
{\it Talk presented at the}\\
{\it Math. Phys. Kiev 1997 International Workshop}\\
{\it in honour of Walter Thirring (Kiev, May 1997)}\\
{\it and also dedicated to the memory of}\\
{\it Volodya Gribov}\\
\end{center}
\vspace*{2.5cm}

\begin{flushleft}
CERN-TH/98-23 \\
January 1998
\end{flushleft}

\vfill\eject

\setcounter{page}{1}
\pagestyle{plain}

\section{Introduction}

In this talk, I would like to speak of the problem of the stability of
three- and four-body non-relativistic purely Coulombic systems. A system
will be said to be stable if its energy is lower than the energy of any
subdivision in subsystems. This is a restrictive definition of stability,
because besides that there are other useful notions: ``metastability" and
``quasi-stability" on which we shall say only a few words later.

The works I will present are
due, in what concerns the three-body case, to J.-M. Richard, T.T. Wu and
myself. The four-body work is due to J.-M. Richard in collaboration with
various other persons (including J. Fr\"ohlich !)

        The reasons why I decided to choose this subject for this workshop
dedicated to Walter Thirring are that I know that Walter is interested in
that topic and that the tools which are used are found precisely in
Walter's celebrated quantum mechanics course: concavity, scaling, and the
Feynman-Hellmann theorem. One tries to avoid, as much as possible,
numerical calculations, or to use already existing numerical calculations
in particular cases. I will speak of:
\begin{itemize}
\item[i)] three-body systems with equal absolute value of the charge,
i.e., $-e +e +e$, or $+e -e -e$, since it is clear that $+e +e +e$ is
unbound. Then binding or no binding will depend on the masses;
\item[ii)] three-body systems with unequal charges;
\item[iii)] four-body systems with charges with equal absolute value. It
will be mostly $+e +e -e -e$. However, I shall say a word on $+e -e -e -e$.
\end{itemize}

\section{Three-body case: equal $\vert$charges$\vert$}

The problem we discuss now is whether a system of three charged particles
(1,2,3), 1 having charge $+e$ and 2 and 3 charges $-e$, is stable or will
dissociate into a two-body system and an isolated particle, (1,2)+3 or
(1,3)+2. The system will be stable if the algebraic binding energy of the
(123) system is strictly less than the binding energy of both (1,2) and
(1,3). If, on the other hand, the infinimum of the spectrum of the (123)
system coincides with the lowest of the (12) and (13) binding energies the
system will be unstable.

This is an old problem which has been treated in many particular cases. For
instance, long ago,  Bethe has shown that the hydrogen negative ion $(p
e^-e^-)$ has one bound state \cite{aaa}, and Hill has shown that there  is only
one such bound state with natural parity \cite{bb}, and Drake has also shown
that there exists an unnatural parity state \cite{cc} and finally Grosse and
Pittner \cite{dd} have shown also that this unnatural parity state is
unique. In
what follows we shall treat only the natural parity states, i.e., states such
that $P = (-1)^L$, where $L$ is the total orbital angular momentum (we neglect
spin interactions!). For three particles there is no problem with the Pauli
principle even if two of them are identical fermions, since we can adjust the
spin.

Wheeler \cite{ee} has also shown that the system $e^+e^-e^-$ is bound, and, more
generally, Hill \cite{bb} has shown that any three-body system in which the two
particles with the same sign of the charge have the same mass is stable. This
covers the two previous cases.

As an example of an unstable system (there are many others !), we can give the
proton-electron-negative muon system, for which a heuristic proof was given by
Wightman in his thesis \cite{ff} and a rigorous proof was given by a
collaboration including Walter Thirring himself \cite{ggg}.

Richard, Wu and myself \cite{hh} have tried to organise the results on
stability, and, by using simple properties, save numerical calculations.
From the reactions we had from experts on numerical calculations we believe
that this was not totally useless.  The three-body Schr\"odinger equation
reads
\beq
-{1\over 2m_1} \Delta_1 \psi -{1\over 2m_2} \Delta_2 \psi -{1\over 2m_3}
\Delta_3\psi
+ \left[ -{e^2\over r_{12}} - {e^2\over r_{13}} + {e^2\over r_{23}}\right]
\psi = E \psi
\label{one}
\eeq
and the corresponding two-body equations can be obtained by omitting some
terms. It is obvious that we have $\underline{\rm scaling}$ properties:
\begin{itemize}
\item[i)] the charges can be multiplied by some arbitrary number without
changing the stability problem;
\item[ii)] the masses can also be multiplied by an arbitrary number, so
that the stability problem depends only on the $\underline{\rm ratio}$ of
the masses, i.e., of $\underline{2}$ parameters.
\end{itemize}

It will be convenient to introduce some variables:
\begin{itemize}
\item[-] the inverse of the masses
\beq
x_1 = {1\over m_1}~~~~x_2 = {1\over m_2}~~~~ x_3 = {1\over m_3}
\label{two}
\eeq
then the ground state energy of the system will be concave in $x_1, x_2,
x_3$, and, in particular, concave in $x_1$ when $x_2$ and $x_3$ are fixed
(and circular permutations!);
\item[-] the $\underline{\rm constrained}$ inverse of the masses
\beq
\alpha_1 = {x_1\over x_1+x_2+x_3}~~~~{\rm etc.}
\label{three}
\eeq
such that
\beq
\alpha_1+\alpha_2+\alpha_3 = 1
\label{four}
\eeq
\end{itemize}
With these new variables, any system of three particles can be represented
by a point in a triangle, $\alpha_1, \alpha_2, \alpha_3$ being the distances
to the sides of the triangle. Figure 1 represents such a triangle with a
few points representing some three-body systems.

\begin{figure}[H]
\hglue5.5cm
\epsfig{figure=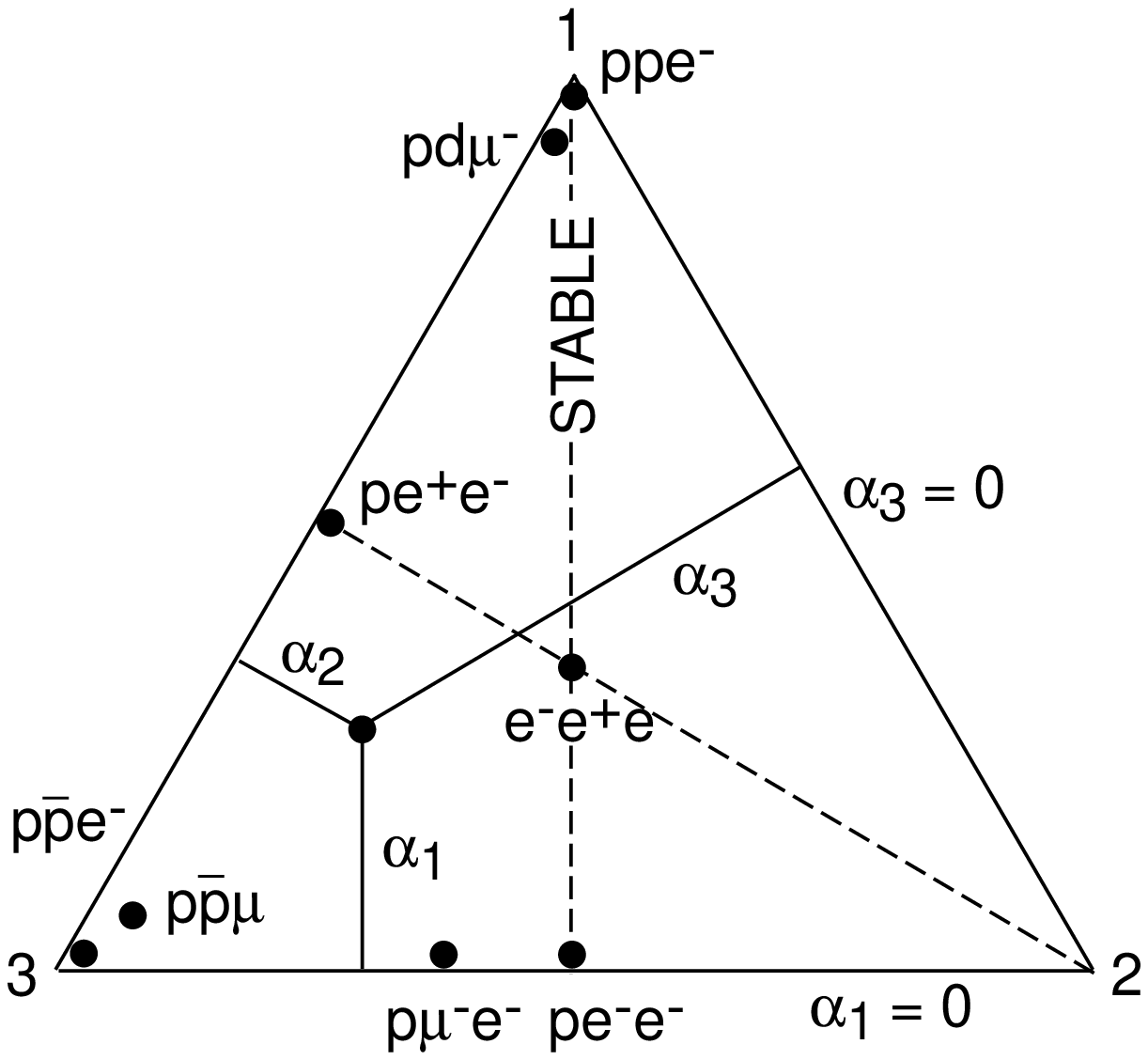,width=5.5cm}
\caption{}
\end{figure}

It is of course sufficient, for the time being (i.e., for $\underline{\rm
equal}$ charges of 2 and 3) to consider the left half of the triangle,
i.e., to assume $m_2 \geq m_3$. Let us remember that since we have,
according to Hill's theorem, strict stability for $m_2 = m_3$, i.e.,
$\alpha_2 = \alpha_3$, there will be some neighbourhood of the line
$\alpha_2 = \alpha_3$ where we shall have stability. However, not all
systems will be stable. We have already mentioned the $p e^-\mu^-$
system as unstable. Another point where instability is obvious is the left
summit marked 3, where we have two infinitely heavy particles with opposite
charge producing zero attraction on the third particle. There is, therefore,
an instability region in the left half triangle.

We have proved three theorems on the instability region in the left half
triangle.

\noindent
{\bf Theorem I} \\ The instability region in the left half triangle is
star-shaped with respect to summit 3.

The proof is based on the Feynman-Hellmann theorem combined with scaling.
take a point $P$ (Fig. 2) where the system is unstable or at the limit of
stability. First we use the variables $x_1, x_2, x_3$. From the
Feynman-Hellmann theorem, ${dE(123)\over dx_3} > 0$, if $x_1$ and $x_2$ are
fixed. The binding energy of the subsystem 12 is fixed. Hence the residual
binding can only increase (algebraically). $x_3$ moves from $x_3(P)$ to
infinity. The image of this in the rescaled $\alpha$ variable is the
segment, $P3$, where $\alpha_1/\alpha_2$ = constant. If there is no binding
at $P$ there is no binding on the whole segment.

\begin{figure}
\hglue5.5cm
\epsfig{figure=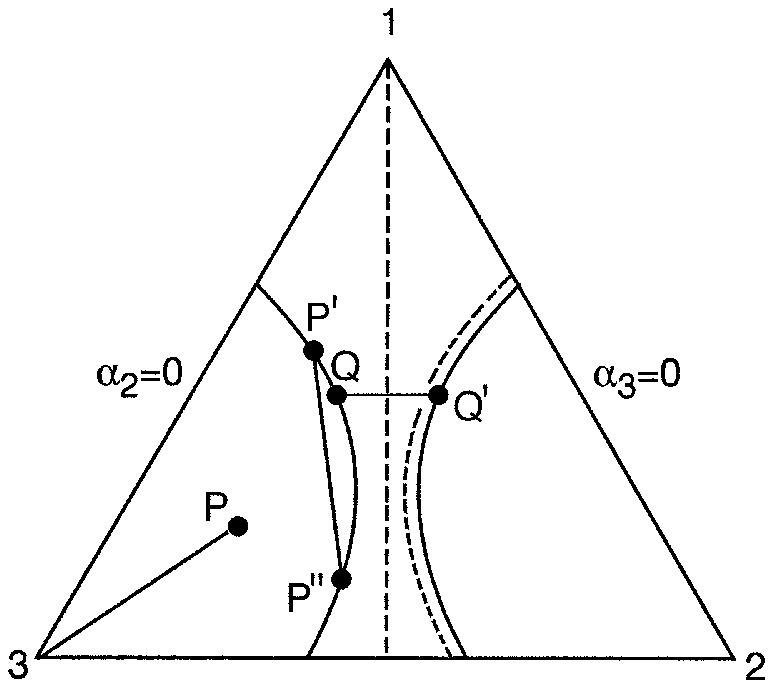,width=5.5cm}
\caption{}
\end{figure}

\noindent
{\bf Theorem II} \\ In the left-half triangle, the instability region is
$\underline{\rm convex}$.

Take two points $P^\prime$ and $P^{\prime\prime}$ on the border of the
stability domain inside the triangle with the $\alpha$ variables. At
$P^\prime$ and $P^{\prime\prime}$ we have $E_{P^\prime}(12) = E_{P^\prime}
(123)$, $E_{P^{\prime\prime}}(12) = E_{P^{\prime\prime}}(123)$. It is
possible to find a linear rescaling $P\rightarrow M$ such that
$E_{M^\prime}(12) = E_{M^{\prime\prime}}(12) = E_{M^\prime}(123) =
E_{M^{\prime\prime}}(123)$. Then one can interpolate linearly between
$M^\prime$ and $M^{\prime\prime}$:
$$
M_\lambda = \lambda M^\prime + (1-\lambda ) M^{\prime\prime}~, \quad\quad 0
< \lambda < 1~~.
$$
For any $M_\lambda$, $E_{M_\lambda}(12)$ = const. and $E_{M_\lambda}(123)$
is $\underline{\rm concave}$ in $\lambda$, and therefore
$E_{M_\lambda}(123) \geq E_{M^\prime}(123) = E_{M^{\prime\prime}}(123)$.
Returning to the original variables $\alpha_1 \alpha_2 \alpha_3$ and
noticing that the scaling is $
\underline{\rm linear}$ we see that, on $P^\prime P^{\prime\prime}$ we have
$E(123) \geq E(12)$. Hence we have instability (Fig. 2).

There is, in fact, a more refined theorem, which we found, following a
question by the late V.N. Gribov during a seminar in Budapest in 1996.

\noindent
{\bf Theorem III} \\ The domain (in the left-half triangle) where
$$
{E(123)\over E(12)} \leq 1 + \epsilon~, \quad\quad \epsilon > 0
$$
is convex.

The meaning of this theorem is that the lines along which the
$\underline{\rm relative}$ binding is constant have a definite convexity.
Note the $\underline{\rm sign}$ of the inequality because $E(123)$ and
 $E(12)$
are both $\underline{\rm negative}$. We believe that the proof is
essentially obvious, since, in the previous theorem, one goes through a
rescaling, replacing $P$ and $P^\prime$ by $M$ and $M^\prime$ where the
two-body energies are equal. Theorem II is of course becoming a special
case of Theorem III, with $\epsilon\rightarrow 0$. The dotted line on Fig. 2
corresponds to some positive value of $\epsilon$.

Let us give a very simple application of Theorem II. We know that,
according to Glaser et al., the system $p_\infty A^-e^-$ is unstable if
$m_{A^-} > 1.57 m_{e^-}$ ($p_\infty$ means a proton with infinite mass).
Similarly, we know that, from the work of Armour and Schrader \cite{jj} the
system $p_\infty A^+B^-$ is unstable if $m_{A^+}/m_{B^-} < 1.51$.

This means that $p_\infty e^-e^+$ is unstable (not because of annihilation
that we neglect, but of dissociation into $p_\infty e^-$ and $e^+$). In
Fig. 3, $p_\infty A^+B^-$ and $p_\infty A^-e^-$ with the limit masses
correspond respectively to $X$ and $Y$. Any point to the left of the
segment $XY$ corresponds, according to Theorem II to an unstable system.
Therefore, $p e^+e^-$, $p \mu^+\mu^-$, with the $\underline{\rm
actual~mass}$ of the proton, are unstable, and one can go up to $p z^-z^+$
which will be unstable if $m_p/m_z > 2.2$.

\begin{figure}[H]
\hglue5.5cm
\epsfig{figure=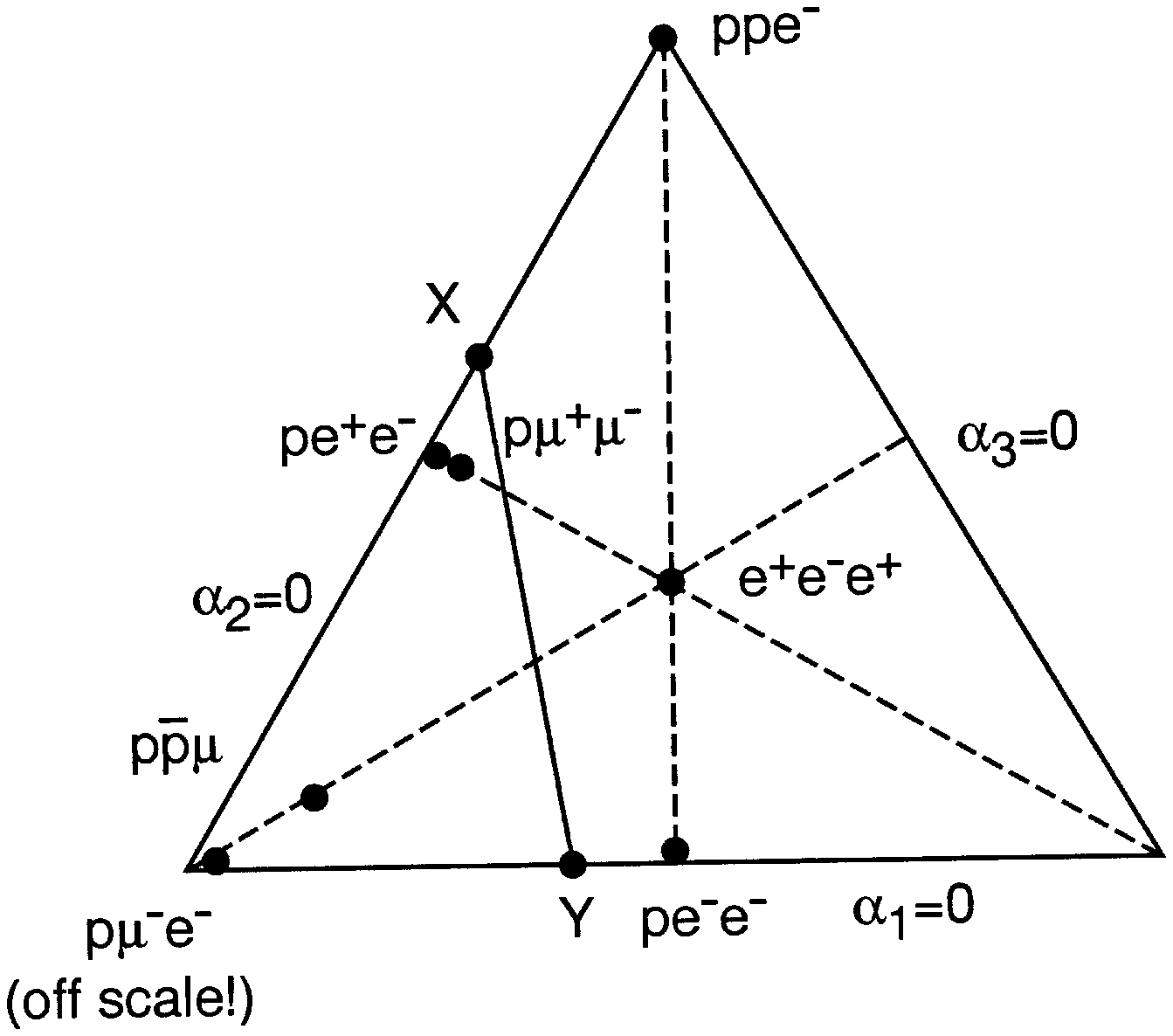,width=5.5cm}
\caption{}
\end{figure}

We obtain too that the system $p \mu^-e^-$, with the actual proton mass is
unstable, and also (disregarding again annihilation) $p\bar p e^-$, $p\bar
p\mu^-$.

One can also use convexity to get results in the opposite direction, i.e.,
prove that certain three-body systems are stable. We know that the systems
represented by a point on the vertical bissector of the triangle are
stable. In practice we know more than that, namely we have an estimate or
more exactly a lower bound of the absolute value of the binding energy of
many systems by using variational calculations and, in fact, by playing
with convexity again it is possible to have a lower bound of the absolute
value of the binding energy at $\underline{\rm any}$ point on the bissector
which corresponds to $\alpha_2 = \alpha_3$, $0 \leq \alpha_2 \leq 1$.

Now we use convexity along a horizontal line $\alpha_1$ = const.

The systems $\alpha_1, \alpha_2, \alpha_3$,~~ $\alpha_1, \alpha_3, \alpha_2$
represented in Fig. 2 by $Q$ and $Q^\prime$, are of course completely
equivalent. Hence
$$
E_{123} (\alpha_1, \alpha_2, \alpha_3) = E_{123}(\alpha_1, \alpha_3, \alpha_2)
< E_{123} \left(\alpha_1, {\alpha_2+\alpha_3\over 2},
{\alpha_2+\alpha_3\over 2}\right)~,
$$
by convexity and
$$
E_{123} \left(\alpha_1, {\alpha_2+\alpha_3\over 2}, {\alpha_2+\alpha_3\over
2}\right) = \bigg(1+g(\alpha_1)\bigg) E_{12}\left(\alpha_1,
{\alpha_2+\alpha_3\over 2}\right) =
\bigg(1+g(\alpha_1)\bigg) E_{12} \left(\alpha_1, {1-\alpha_1\over
2}\right),
$$
where $g$ represents the relative excess in binding energy. We are assured of
stability if
$$
E_{12}(\alpha_1,\alpha_2) > \bigg( 1+g(\alpha_1)\bigg) E_{12} \bigg(
\alpha_1, {1-\alpha_1\over 2}\bigg)
$$
i.e., if
$$
{e^2\over 2}~~{1\over\alpha_1+\alpha_2} < \bigg( 1+g(\alpha_1)\bigg)
{e^2\over 2}~~{2\over 1+\alpha_1}
$$

In this way, it is possible to prove that the system $p d \mu^-$, important
for fusion processes, is stable, through it is off the diagonal. One would
like to show also in this way that $\pi^+\mu^-\mu^+$ and $\mu^+\pi^+\pi^-$
are stable, but these considerations are not sufficient. There is a hint
that they are stable because, from explicit calculations at $\alpha_1 = 0$
one sees that this method tends to give a band of stability which is two
times narrower than the real one and this is just what one needs.

\section{Three-body case. Unequal charges}

On this topic, Richard, Wu and myself have published one paper \cite{kk} and
one in preparation, of which I shall give some of the really new results.
We have the right to take $q_1 = 1$, the charge of the particle which is
opposite to the other two, of the same sign, $q_2$ and $q_3$.

\noindent
{\bf A) Unequal charges, but $q_2 = q_3$}

This is the simplest case, very similar to the case of all equal charges.
For fixed $q_2 = q_3$ we can again represent a system with the variables
$\alpha_1 \alpha_2 \alpha_3$, and, on the bissector of summit 1, the
energies of the subsystems (12) and (13) are equal. The fact that the
instability regions are star-shaped with respect to 3 for the left-half of
the triangle and to 2 for the right-half persists, and as well the
convexity of the instability regions. There are two major differences which
are:
\begin{itemize}
\item[i)] that if $q_2=q_3 < 1$, all three-body systems are stable, because
near summit 3, for instance, the subsystem (12) is very compact and exerts a
Coulomb attraction at long distances on particle 3; it may seem strange
that as $q_2  \rightarrow 1$ part of the triangle becomes unstable, but
this is just due to the fact that the binding energy, in that region, tends
to zero as $q_2 \rightarrow 1$;
\item[ii)] that if $q_2 = q_3$ is large enough, stability disappears
completely.
\end{itemize}

Figure 4 summarizes the situation. For $q_2 > 1$ but very close to 1, there
is no qualitative difference but for a certain critical value $1 < q_{2c} <
1.1$, the stability band breaks into two pieces, and from calculations by
Hill and collaborators \cite{lll} stability near $\alpha_2 = \alpha_3 = 1/2$,
$\alpha_1 = 0$ disappears completely for $q_2 \ge 1.1$, and from the calculations
of Hogr\`eve it disappears near $\alpha_2 = \alpha_3 = 0$, $\alpha_1 = 1$ for
$q_2 > 1.24$. From convexity, it hence disappears $\underline{\rm completely}$
along the segment joining $\alpha_2 = \alpha_3 = 0$ and $\alpha_2 = \alpha_3 =
1/2$, and from the star-shaped property, there is no stability at any point in
the triangle for $q_2 > 1.24$.

\begin{figure}[H]
\hglue4.5cm
\epsfig{figure=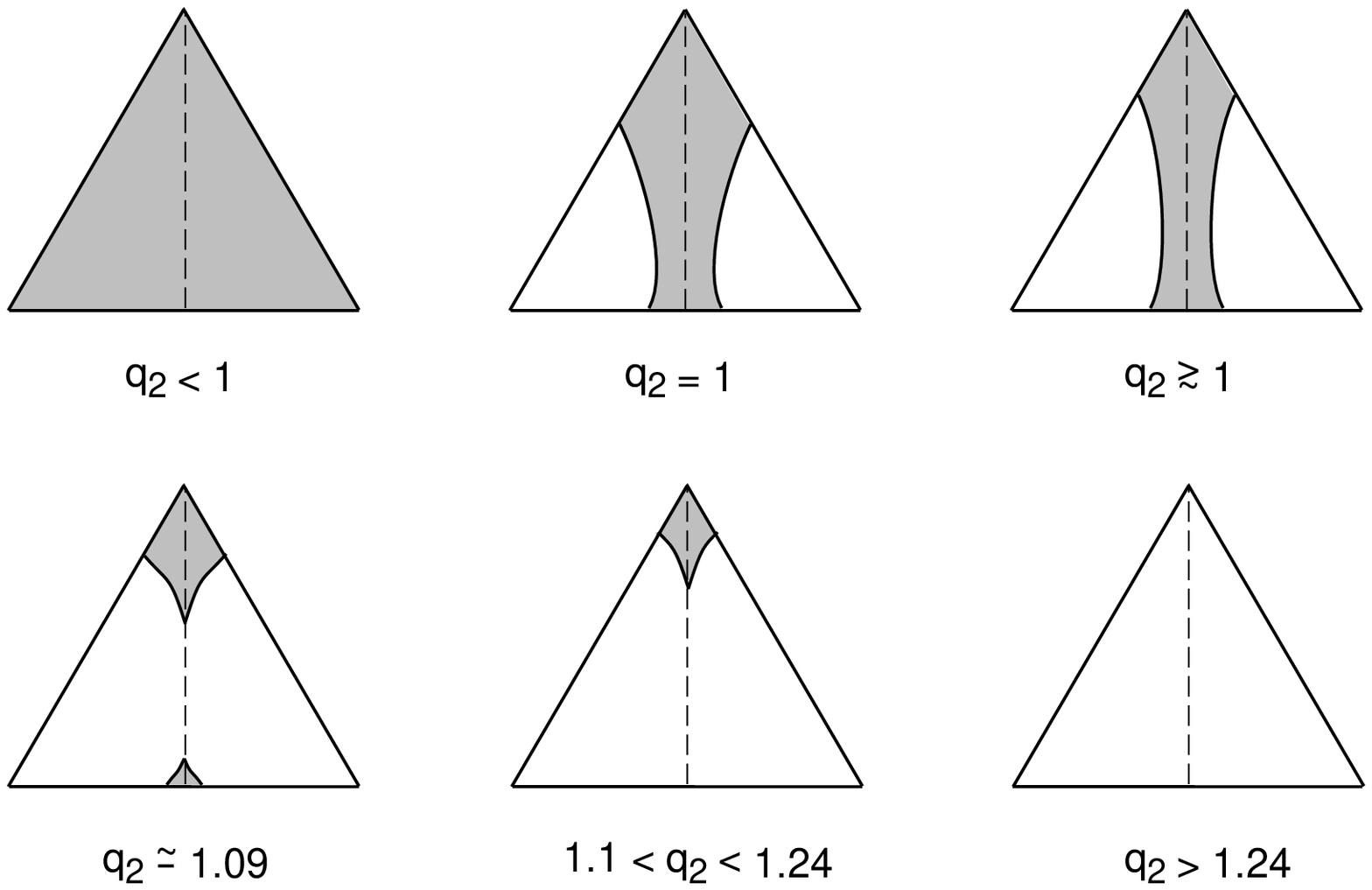,width=8.5cm}
\caption{}
\end{figure}

\noindent
{\bf B) Unequal charges, but $q_2 \not= q_3$ fixed}

First we continue to use the $\alpha_1,  \alpha_2, \alpha_3$
variables to describe the three-body system for fixed charges. A
fundamental difference is that the bissector of summit 1 of the triangle
no longer plays a special role. It is, instead, the line along which
$E_{12} = E_{13}$, i.e.,
$$
{q^2_2\over\alpha_1+\alpha_2} = {q^2_3\over \alpha_1+\alpha_3} \quad {\rm
or} \quad q^2_2 \bigg( 1 - \alpha_2\bigg) = q^2_3 \bigg( 1 -
\alpha_3\bigg)~,
$$
which becomes important. This line goes through the point $\alpha_2=\alpha_3
= 1$, symmetric of summit 1 with respect to the line $\alpha_1 = 0$ (Fig.
5). The line divides the triangle into two subregions. If we decide to take
$q_2 \geq q_3$, $E_{12} < E_{13}$ in the left region which contains the
summit 1.

\begin{figure}[H]
\hglue6.5cm
\epsfig{figure=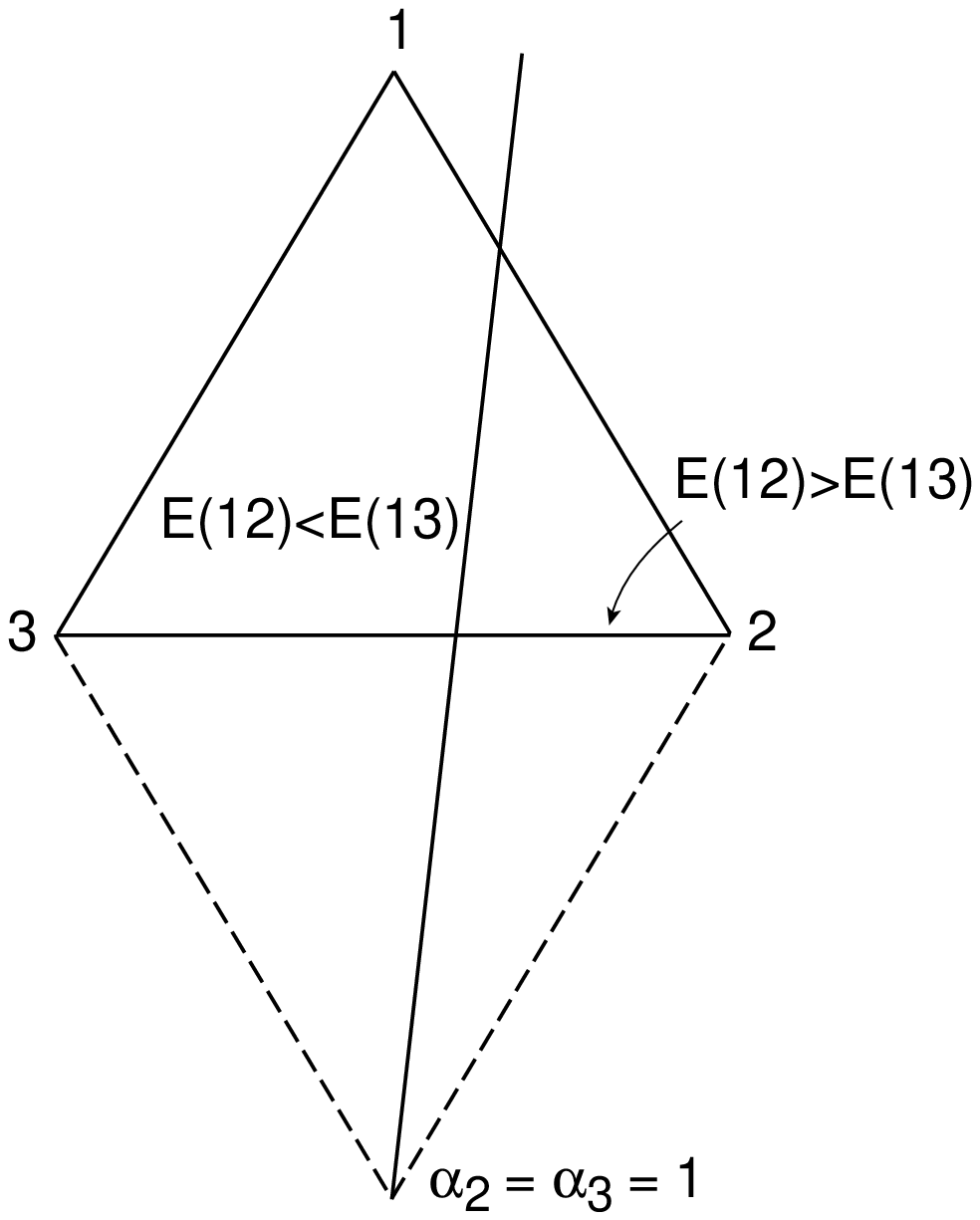,width=4.5cm}
\caption{}
\end{figure}

If $q_2$ and $q_3$ are both less than 1, we have again stability
everywhere. If $q_2 \geq 1$ with $q_3 < 1$, part of the triangle becomes
unstable. Various scenarios are shown in Fig. 6. Notice that summit 3, on
the left is unstable, together with some neighbourhood. This is because
$\alpha_1 \mathrel{\rlap{\raise.5ex\hbox{$\sim$}}
{\lower.5ex\hbox{$=$}}}
\alpha_2 \mathrel{\rlap{\raise.5ex\hbox{$\sim$}}
{\lower.5ex\hbox{$=$}}} 0$ corresponds to a very compact system with either
very weak rapidly decreasing attraction on particle 3 (if $q_2 = 1$) or
repulsion (if $q_2 > 1$).

\begin{figure}
\hglue4.5cm
\epsfig{figure=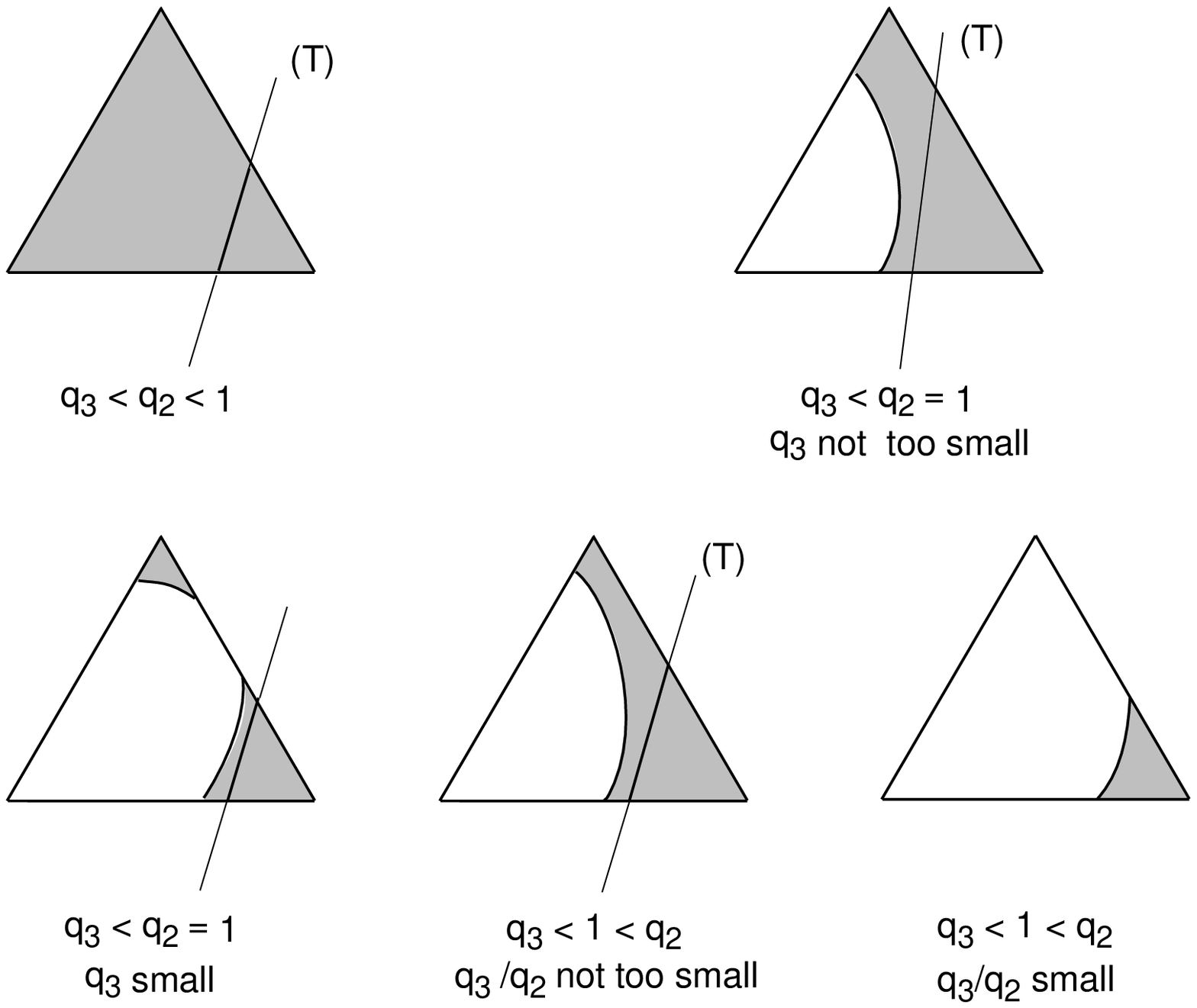,width=8cm}
\caption{}
\end{figure}

For $q_2$ and $q_3$ both $>1$, we have only a partial understanding of the
situation. If they are both not too large, a neighbourhood of the dividing
line will survive as stability region plus a neighbourhood of summit 1.
Summits 2 and 3 will definitely be unstable.

We believe that if $q_2$ and $q_3$ are sufficiently large there is no
stability at all, for any mass, but this is very difficult to implement
quantitatively except in two places:
\begin{enumerate}
\item
at summit 1, which corresponds to $\alpha_2 = \alpha_3 = 0$, and is the
Born-Oppenheimer limit, for which Hogr\`eve \cite{mm} has shown that one
has instability if either $q_2$ or $q_3 > 1.24$;
\item at the point $\alpha_1 = 0$, $\alpha_2 = \alpha_3 = 1/2$ where Lieb's
theorem \cite{nn} applies: if $1/q_2 + 1/q_3 \leq 1$ there is no stability.
\end{enumerate}

This implies by convexity that if $q_2$ and $q_3 > 2$ one has instability
for $\alpha_2 = \alpha_3$, and by the star property for $\alpha_2 <
\alpha_3$ (if $q_2 > q_3$). For a considerable improvement, see D.

\noindent
{\bf C) $q_2$ and $q_3$ variable, fixed masses}

Instead of holding charges fixed one can fix the masses and study stability
in the $q_2,  q_3$ plane. One particular case is $m_2 = m_3 = \infty$ where
one has the Born-Oppenheimer limit and one has the diagram calculated by
Hogr\`eve \cite{mm} (Fig. 7).

\begin{figure}[H]
\hglue4.5cm
\epsfig{figure=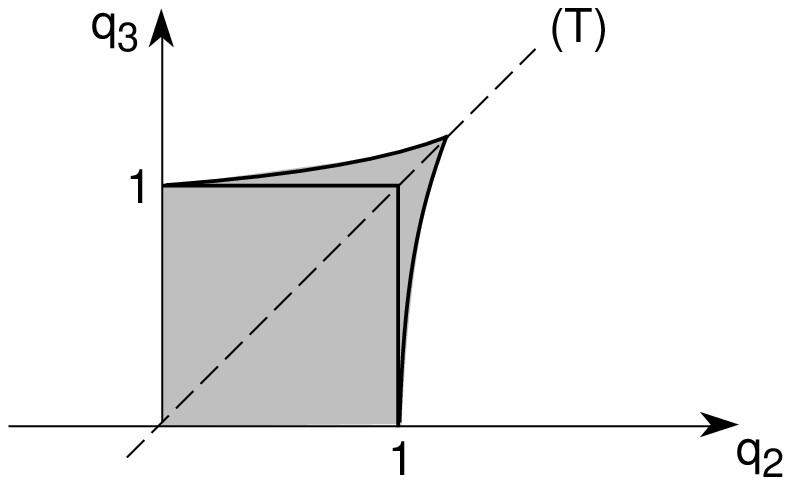,width=7.5cm}
\caption{}
\end{figure}

In the $q_2,  q_3$ plane there is again for the general mass case a dividing line where
the binding energies of the two subsystems  (12) and (13) are equal:
$$
q^2_2 {m_2 \over m_1 + m_2} = q^2_3 {m_3\over m_1 + m_3}~.
$$
In the two sectors thus defined there are two instability regions for which
we have been able to derive a new concavity property:

\noindent
{\bf Theorem IV}

Define $z_2 = 1/q_2, z_3 = 1/q_3$, the image of $q_2 > 0~q_3 > 0$ is $z_2 >
0~z_3 > 0$. Then, in the $z$ variables the two instability regions are
$\underline{\rm convex}$ (Fig. 8). The proof is based on a rescaling such
that the binding energy of the relevant subsystem remains constant on a
segment in the $z_2 z_3$ plane.

\begin{figure}[H]
\hglue5.5cm
\epsfig{figure=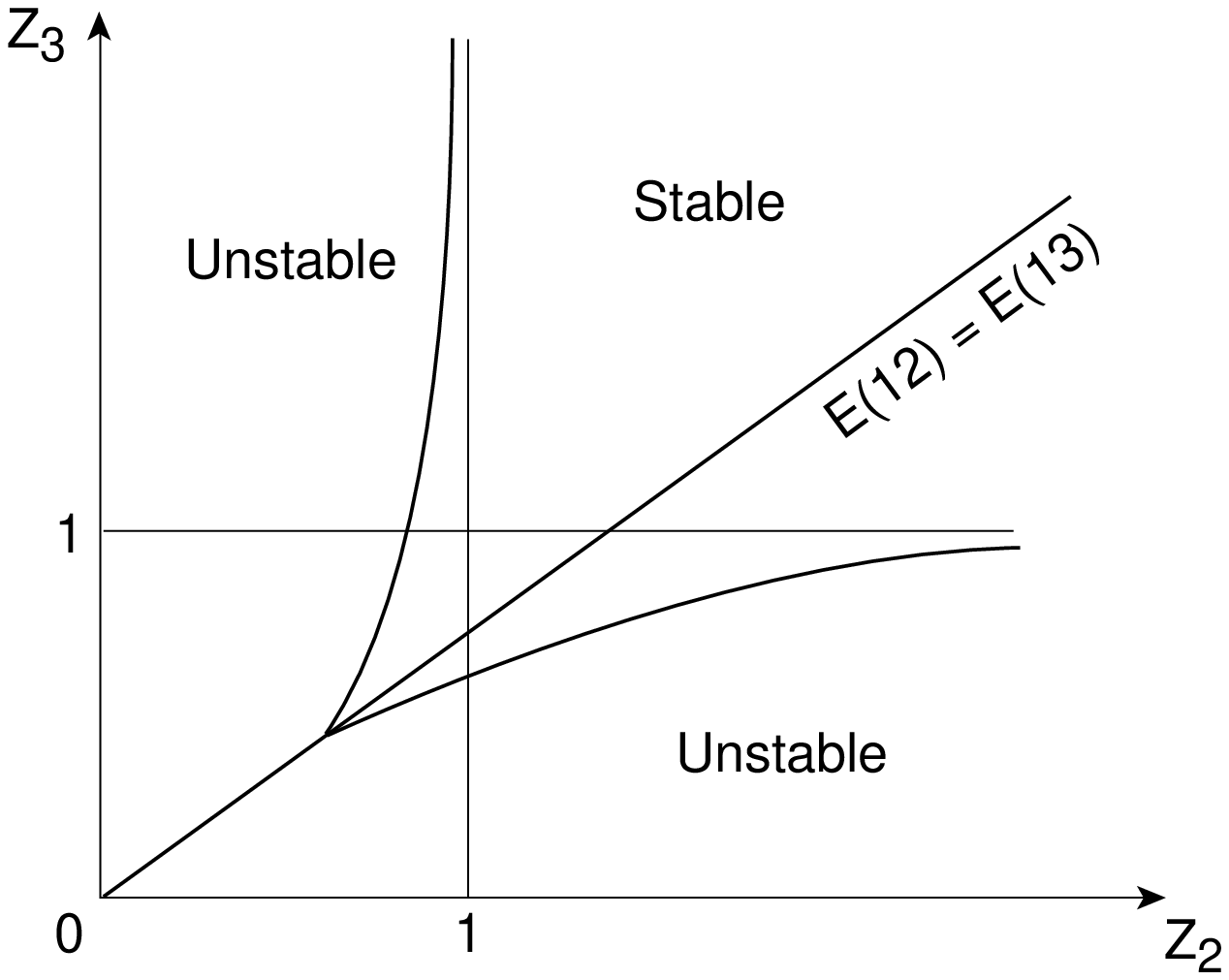,width=5.5cm}
\caption{}
\end{figure}

\noindent
{\bf D) An illustration: the instability of the systems $\alpha p e^-$ or
$\alpha p \mu^-$}

In the Born-Oppenheimer limit it is known that such systems are unstable
\cite{oo}. Spruch and collaborators \cite{pp} have given arguments which
seem to indicate that this might remain true for the actual masses of the
protons and of the $\alpha$ particle, but, to our knowledge, a completely
rigorous proof does not exist.

\noindent
{\bf STEP 1} ~~ Take $m_2 = m_3$. Then if $q_2 = q_3$ we have instability if
$q_2 = q_3 \geq 1.24$ and $m_2 = m_3 = 0$ and $m_2 = m_3 = \infty$, and by
concavity for any $m_2 = m_3$. Now consider the segment $q_2$ = 1.24, $0 <
q_3 < 1.24$. Along this segment the subsystem with the most negative
binding energy is (12) and this energy is constant. There is no stability
for $q_2 = q_3 = 1.24$, and no stability for $q_2 = 1.24$, $q_3$
very close to zero, because then particle 3 is submitted to a very weak
force and is therefore very far away most of the time while 12 is overall
repulsive for 3. By concavity there is no stability for $q_2 = 1.24$,  $0 <
q_3 < 1.24$. The same kind of argument applies to $1.24 < q_3 < \infty$,
because one has instability for $q_3 \rightarrow \infty~~ q_2 > 1$, and 
one can use
concavity in the inverse charge. The conclusion is that if $m_2 = m_3$, 
one has no
stability if either
$q_2
\geq 1.24$ or $q_3
\geq 1.24$.

\noindent
{\bf STEP 2} ~~ Assume $q_2 \geq q_3$. Then,  in the whole sector, $m_2 >
m_3$, the lowest two-body threshold is given by the (12)  system. therefore
in the $\alpha$ triangle the region $\alpha_2 < \alpha_3$, is completely
unstable for $q_2 > 1.24$ by use of the star shaped property.

The systems
$$
\matrix{ \alpha p e^- & \alpha p \mu^- \cr
\alpha d e^- & \alpha d \mu^- \cr
\alpha t e^- & \alpha t \mu^-}
$$
satisfy precisely the conditions: $q_2 > 1.24,~~ q_3 = 1,~~   m_2 > m_3$,
and are therefore unstable, in the sense we have given to ``instability".
$\underline{\rm Notice ~that ~the~ proof~ would~ fail ~if~ the ~particle~
with }$
$\underline{\rm  charge~2~ was ~lighter ~than~ the~ particle~ with~ charge~
1.}$

However, as pointed out by for instance Gerstein \cite{qq}, some of the
levels of these systems are ``quasi stable" in the Born-Oppenheimer
approximation in the sense that the minimum of the Born-Oppenheimer
potential is below the value it takes for infinite separation between the
two nuclei where one of the limit atomic states is excited and degenerate with
the other one. Gerstein \cite{rr}
went as far as estimating the lifetimes of these quasi-stable states and
showed that the lifetime increases drastically when the proton is replaced by a triton.
One should also mention the metastability, where the Born-Oppenheimer curve has a
minimum above zero \cite{mm}.

\section{Four-body case: equal charges}

Most of what I will say concerns systems of two positive and two negative
charges of absolute value $e$.

However, let me start with the case
$$
p e^-e^-e^-~,
$$
i.e., a doubly negative hydrogen ion.

Such a state according to a review  by Hogr\`eve \cite{ss} does not seem to
exist. In the limit of an infinitely heavy proton, the Lieb bound on $n$, the number
of electrons around a charge $Z$
\cite{nn},
$n < 2Z + 1$, which is a strict inequality, gives $n < 3$. In fact no
doubly negative atomic ions seem to exist in nature, while singly negative
ions may (like $H^-$) or may $\underline{\rm not}$ exist (like the case of
the rare gases).

We return now to systems with charges $-e -e +e +e$, and first of all $m^-
m^- M^+ M^+$, i.e., two negatively charged particles with equal mass and
two positively charged particles with equal masses. A familiar example is
the hydrogen molecule $e^-e^-p^+p^+$. A more exotic example is the
positronium molecule $e^-e^-e^+e^+$.

It has been realized by Jurg Fr\"ohlich that up to very recently there did
not exist any rigorous proof of the stability of the hydrogen molecule. It
was believed to be stable because of experiment of course, and of
Born-Oppenheimer calculations.

Two groups (Fr\"ohlich et al., Richard) investigated this problem and
finally joined their efforts to produce a completely rigorous proof
\cite{tt}.

The simplest approach, whose idea comes from J.-M. Richard \cite{uu}
consists of  starting from the work of \O re, which is valid by scaling for a system
$A^- A^- A^+ A^+$ \cite{vv}.  \O re used a very simple variational trial function, of
the form
$$
\matrix{
\psi = &\exp -{1\over 2} \bigg( r_{13} + r_{14} + r_{23} + r_{24} \bigg) \cr
&\cr
&\cosh \bigg[ {\beta \over 2} \bigg( r_{13} - r_{14} - r_{23} + r_{24}
\bigg) \bigg ]}
$$

Notice that the distances between particles with same charge sign do not
appear. All integrals can be carried analytically and it is found that the
energy is less than
$$
2.0168 ~~~ E_0 (A^+A^-)~.
$$

The system is therefore stable because it cannot dissociate into $A^+A^- +
A^+A^+$. It cannot dissociate either in $A^+A^-A^+ + A^-$, because between
these two systems there is a long-distance Coulomb force, producing
unavoidably infinitely many bound states.

If we take now
$$
x_e + x_p = {1\over m_e} + {1\over m_p} = {2\over m_A}~,
$$
we see that the binding energy of $e^-p$ is the same as that of $A^+A^-$.
However,
$$
E (x_e, x_e, x_p, x_p) = E(x_p, x_p, x_e, x_e) <
E\bigg( {x_p+x_e\over 2}, {x_p+x_e\over 2}, {x_p+x_e\over 2}, {x_p+x_e\over
2}\bigg)~,
$$
by concavity in the inverse masses.

So,
$$
E(p^+, p^+, e^-, e^-) < E (A^+,A^+, A^-, A^-) < 2E(A^+A^-)
$$
Hence, $ppee$ is stable.

One can wonder if stability remains if the masses of two particles of the
same charge are different, i.e.,
$$
A^+B^+C^-C^-
$$

Then there is still a unique possible dissociation threshold:
$$
A^+C^- + B^+C^-
$$
\O re has predicted explicitly that the system $p e^+e^-e^-$ is
$\underline{\rm stable}$ \cite{ww}, and this has been observed
experimentally by Schr\"ader and collaborators \cite{yy}.

It is also easy, from the upper bound of the energy of $e^+e^+e^-e^-$ to
show that the systems $p~ d~ e^-e^-$, $p~ t~ e^-e^-$, $d~ t~ e^-e^-$ are stable. This
is implicit in the work of Richard \cite{uu}, established in the thesis of
Seifert \cite{zz} and I present here my own version. By concavity we have
$$
\matrix{
&E(x_A,x_B,x_C,x_C) < E\bigg( {x_A+x_B\over 2}, {x_A+x_B\over 2}, x_C,
x_C\bigg) \hfill\cr &\cr
&< E\bigg( {x_A+x_B\over 4} + {x_C\over 2}, {x_A+x_B\over 4} + {x_C\over
2}, {x_A+x_B\over 4} + {x_C\over 2}, {x_A+x_B\over 4} + {x_c\over 2}\bigg)
\hfill\cr &\cr
&< -2.0168 ~~{1\over 4} ~~{1\over {x_A+x_B\over 4} + {x_C\over 2}}\hfill }
$$
If the inequality
$$
-2.0168 {1\over {x_A+x_B\over 4} + {x_C\over 2}} < - \left( {1\over
{x_A\over 2} + {x_C\over 2}} + {1\over {x_B\over 2} + {x_C\over 2}}
\right)~,
$$
is satisfied, the system is stable. If $m_A > m_B > m_C$ one finds that
this condition is certainly satisfied if $m_B > 5m_C$.

Using the more refined bound \cite{tt}
$$
E_{A^+A^+A^-A^-} < -2.06392 E (A^+A^+)~,
$$
which uses more sophisticated trial function and must be ``cleaned" from
numerical roundup errors, one gets
$$
m_B > 2.45 m_C~.
$$

However, Varga and collaborators \cite{aai}, using
trial functions leading to integrals which can be expressed analytically,
and adjusting parameters, have found that one has stability for any $m_A$
and $m_B$, including the case where one or two of them are less than $m_C$.
By a tedious but feasible exercize, one could, using concavity, transform
this calculation, which is unavoidably done for discrete values of the
masses, into a very inelegant proof. Let us hope that someone, in the
future, will find a still more clever trial function and avoid this.

Let us mention, finally, a conjecture by Jean-Marc Richard: ``A four-body
system with two positive-charge and two negative-charge particles (equal in
absolute value) is stable if one three-body subsystem is
stable".

\end{document}